\documentstyle[prb,aps,twocolumn]{revtex} 
\begin{document}
\title{An unusual interplay among disorder, Kondo-effect and spin-glass
behavior in the Kondo lattices, Ce$_2$Au$_{1-x}$Co$_x$Si$_3$}

\author{Subham Majumdar and E.V. Sampathkumaran$^*$}

\address{Tata Institute of Fundamental Research, Homi Bhabha Road, 
Colaba, Mumbai - 400005, India\\
$^*$E-mail address: sampath@tifr.res.in}

\author{St. Berger, M. Della Mea, H. Michor and E. Bauer}

\address{Institut f.Experimentalphysik, Vienna University of Technology,
A-1040 Wien, Austria}

\author{M. Brando, J. Hemberger and A. Loidl}

\address{Experimentalphysik V, Elektronische Korrelationen und
Magnetismus, Institut f\"ur Physik, Universit\"at Augsburg, D-86135
Augsburg, Germany}

\maketitle

\begin{abstract} {{\bf We report the results of magnetic measurements
for the solid solution Ce$_2$Au$_{1-x}$Co$_x$Si$_3$. The results
reveal that this solid solution is characterized by a magnetic phase
diagram (plot of magnetic transition temperature versus $x$) unusual for
Kondo lattices.  In particular, the spin-glass freezing induced by disorder 
is observed only
for the compositions at the weak coupling limit; 
as one approaches 
the quantum critical point by a gradual 
replacement of Au by Co, this disorder effect is surprisingly suppressed in favor of 
long range antiferro-magnetic
ordering in contrast to expectations. 
This unusual interplay between disorder, spin-glass
freezing and the Kondo-effect calls for further refinement of theories on
competition between magnetism and the Kondo effect.}}

\end{abstract}
\maketitle
\vskip0.5cm

The study of the consequences of the competition between the Kondo effect
and magnetic ordering in rare-earth and actinide systems continues to be an active topic
of research. While the observation of long range magnetic ordering in some
 Kondo lattices itself is a subject of great theoretical interest
considering that the Kondo effect should quench magnetism, one also
observes disorder-induced spin-glass anomalies in the vicinity of quantum
critical point (QCP) in some U and Ce systems (see, for instance, Refs. 1
and 2). This made this topic of research more fascinating and new
theoretical ideas have been initiated in the recent literature under the
assumption that such disorder effects are operative only at
QCP.\cite{3,4,5} In this Letter, we bring out a counter-example, $viz.,$
the recently discovered\cite{6,7} solid solution,
Ce$_2$Au$_{1-x}$Co$_x$Si$_3$, in which spin-glass behavior in fact is
observed only when one approaches the weak coupling limit, that is, as one 
moves towards
larger unit-cell volume. In otherwords, this glassy behavior
surprisingly gets suppressed with increasing chemically-induced disorder 
as one approaches QCP, as though increasing Kondo interaction
strength plays an unusual role of favoring long-range magnetic order at
the expense of spin-glass behavior. This finding adds a new dimension to
this direction of research, demanding further refinement of the theories
on this subject.

These alloys, crystallizing in a AlB$_2$-derived hexagonal structure, have
been found to exhibit many interesting magnetic characteristics.\cite{7}
While the parent Au compound orders magnetically below about 3 K,
Ce$_2$CoSi$_3$ is a non-magnetic Kondo lattice. As Au is progressively
replaced by Co, the magnetic ordering temperature (T$_o$) goes through a
maximum (to about 8 K) as a function of $x$. This observation indicated
that this compound lies at the left hand side of the Doniach's magnetic
phase diagram.\cite{8} This alloy series is one of the few solid solutions
spanning a wide range of Doniach's magnetic phase diagram\cite{8} thus
serving an ideal series to test a theory for a large variation in the 4f
exchange interaction strength.  It is also to be noted that, for some
intermediate compositions before the maximum, two magnetic transitions
were observed with the previous data revealing interesting changes in
magnetic structure both as a function of $x$ and T. This prompted us to
investigate exact nature of these transitions in all these alloys to unravel the
mysteries behind the competition between the Kondo effect and magnetism.  We
have therefore performed ac susceptibility ($\chi_{ac}$) measurements on
many compositions of this alloy series at low temperatures. In addition,
 in order to augment the line of arguments of these alloys, we carried out
low temperature isothermal magnetization (M) (in addition to that reported in
Ref. 7) and low temperature heat-capacity (C(T)) of
Ce$_2$Au$_{0.2}$Co$_{0.8}$Si$_3$. The knowledge thus gained viewed
together with the findings reported in Ref. 7 enables us to construct a
magnetic phase diagram.

The samples employed were synthesized as discussed in Ref. 7. The x ray
diffraction patterns reveal that all the samples are single phases, though
there could be a small amount of an impurity phase for x= 0.0 if
interpreted within AlB$_2$ structure. The $x$= 0.0 sample is a
single phase, if the pattern is analysed in an orthorhombic
structure.\cite{7}  For many alloys ($x$= 0.0,
0.1, 0.2, 0.3, 0.4, 0.6 and 0.8), $\chi_{ac}$ (1.8-15 K) was probed
employing a commercial magnetometer at various frequencies (1, 10, 100,
1000 Hz).  Isothermal dc M data were also collected   
below 12 K for some compositions at the Au rich end with the same 
magnetometers. We have also taken C data down to 1.2 K to resolve the
question\cite{7} whether the composition $x$= 0.8 close to the QCP is a
spin-glass.

In Fig. 1, we show $\chi_{ac}$ data, both for the real ($\chi \prime $)
and the imaginary ($\chi \prime \prime $) components, as a function of
(logarithmic) T below 12 K for $x$= 0.0. There is a peak at about 3 K due
to the onset of magnetic ordering consistent with the observation in the C
and dc $\chi$ data\cite{7}.  There is an additional peak at about 2.2 K
(which could not be resolved in earlier studies\cite{6,7} due to
non-accessibility of this T range), establishing the existence of another
magnetic transition. In fact two transitions could in principle be
expected\cite{9} from two crystallographically inequivalent Ce sites for
an ordered arrangement of Au and Si ions, for which there is an evidence
from the doubling of $a$ parameter of the unit-cell.\cite{7} Therefore, at
present, we cannot resolve whether the double transition arises from these
two sites or whether the same Ce ions undergo spin-reorientation as T is
lowered from 3 to 2.2 K.  Our main conclusions are however not affected by
actual reasoning of these two transitions. 

We now address the question whether the disordered alloys are spin-glasses 
and if so whether this changes with composition. 
For this purpose, we show in Fig. 1 the
frequency dependence of $\chi_{ac}$ for several compositions ordering
magnetically in the covered temperature range.  It is a well known fact
that the peak of $\chi_{ac}$ should not exhibit a frequency dependence for
long range ordering at the transition temperature,\cite{10} but spin
glasses, however, are characterized by a peak shift of the order of a few
percent. This essentially means a shift of the peak by about 0.1 K in our
alloys for a change of frequency from 1 Hz to 1 kHz, which can easily be
missed if the peak is broad as it appears to be the case in the present
series.  However, we can draw an inference on the frequency dependence if
the left part of the curve moves to a higher temperature with increasing
frequency. Also, in spin glasses, $\chi \prime \prime$ shows a sharp and
prominent anomaly near the transition.\cite{10} If one looks for these two
features, it follows that both the transitions for $x$= 0.0 interestingly
could be due to spin-glass freezing.  It is therefore clear that the broad
C(T) anomaly spreading over a wide temperature range\cite{7} is associated
with such a disordered magnetism.

With respect to the $x$ = 0.1 and 0.2 compositions, magnetic transitions 
are observed in the vicinity of 7-8 K as well as close to
4.5~K, manifesting as a broad shoulder and a peak (or a sharp upturn) in
$\chi \prime$ respectively; the 4.5~K-transition is of a spin-glass-type
within the scope of the criteria stated above, but the 8~K-transition is
not. This interpretation is consistent with the ZFC-FC dc $\chi$ behavior
reported earlier.\cite{7} Incidentally, there is also a prominent peak in
C(T) (see Fig. 3 of Ref. 7) around 7 K confirming the existence of another
magnetic transition around this temperature.  Correspondingly, inverse dc
$\chi$(T) also exhibits a flattenning below this temperature (see Fig. 4
inset in Ref. 7).  The fact that the 7K-transition arises from magnetic 
ordering only is established further from   the 
isothermal M behavior  discussed (in figures 6 and 7) in Ref. 7. We have taken 
additional M data at various temperatures at the Au rich end to strengthen 
our conclusions.  The behavior of M for x= 0.1 below 15 K is  
shown in Fig. 2a to highlight the following point: while the plot is linear 
at 10 K typical of paramagnets, there is a sharp
upturn of M beyond 2.5 kOe at 5 K.
This feature becomes broader as one increases temperature, say, to 6.5 K. However,
such a feature is absent at 2 K.  The behavior is the same for x= 0.2.
These observations suggest that there is a metamagnetic transition in a narrow 
temperature interval above 4.5 K. This establishes that the zero-field magnetic 
ordering in this narrow temperature range is of an antiferromagnetic type. We may add that  
the observation of such spin-reorientation effects in M data is not characteristic of 
Schottky anomaly, but only of antiferromagnetism, thereby conclusively ruling out
any explanation of the 7K-peak in the C data in terms of
Schottky anomalies. It may be recalled\cite{7} that, the isothermal M behavior at 5 K
for $x$= 0.0 does not exhibit such an effect, thereby ruling out antiferromagnetic state
for this end member. 

Spin glass features are seen for $x$= 0.3 with a transition close to 5.3
K (with another feature at 2.7 K). It appears that the 7K-transition seen
for x= 0.2 has moved down to a lower temperature (close to 6 K) and now it
appears as a tail above the peak in the $\chi \prime $ data.  For $x$=
0.4, however, there is only one magnetic transition at 5 K which nicely
agrees with C(T) and dc $\chi$ data,\cite{7} with a frequency dependence
of the curve below the peak, classifying this composition also as a
spin-glass. Interestingly, for x= 0.6, not only the peak temperature (4.5
K) but also the values are frequency independent throughout the T range of
measurement, as if this alloy cannot be classified as a spin-glass; in
support of this, the $\chi \prime \prime$ anomaly at the magnetic
transition is negligible. Therefore, ZFC-FC dc $\chi$ bifurcation reported
earlier\cite{7} for this composition does not arise from spin glass
freezing and it should be noted that such features have been observed in
the past even in some long range magnetically ordered systems.\cite{11} 

In short, judged by the frequency dependence of $\chi \prime$ below the peak
as well as by distinct $\chi \prime \prime$ features, we conclude that all
the compositions below $x$= 0.6 are spin-glasses at very low temperatures
(typically below 4.5 K). However,
 long range magnetic ordering sets in at slightly higher temperatures for
$x$= 0.1-0.3. 

We now turn to the magnetism of the Co-rich end. 
In order to probe this aspect,  the C measurements were
performed down to 1.2~K for the composition x= 0.8 
and the results are shown in Fig.
2b. It is distinctly clear that there is a prominent peak at 2 K. In
spin-glass systems, the corresponding feature in C is broadened
significantly and the observation of a well-defined peak therefore rules
out spin-glass freezing for this composition (which is close to QCP, given
that the end Co member is non-magnetic).  The $\chi \prime $ data (Fig. 1)
tends to show a flattening below 2 K (without any frequency dependence of
absolute values) as though there is an onset of magnetic order very close
to this T; the absence of $\chi \prime \prime$ feature at 2 K also rules
out spin glass freezing for this composition.

We would like to mention that we have also measured\cite{12} MR (=
[$\rho$(H)-$\rho$(0)]/$\rho$(0)) as a function of H down to 0.6 K
primarily to support our conclusions for the compositions close to QCP
($x$= 0.6 and 0.8).  The MR of these alloys is found to be {\it positive}
(typically close to 2\%) even for reasonably large values of H (say till
40 to 60 kOe) well below their respective T$_o$, characterizing that these
alloys are antiferromagnets without magnetic Brillouin-zone gap effects at
temperatures as low as 0.6 K. If the low temperature state is a
spin-glass, the sign of MR should be negative even in the low field range.  
We also observed a sign reversal at higher fields establishing the
field-induced spin-flipping in these alloys. Above the respective ordering
temperatures, MR is negative in the entire field range typical of
paramagnets.

By consistently interpreting the features observed in all these data as
well as dc $\chi$ and C features reported in Ref. 7, we have sketched a
magnetic phase diagram (see Fig. 3) for the present solid solution. 
For this purpose, the 
onset of spin-glass freezing marked in Fig. 3 is inferred from 
the  temperature at which $\chi \prime \prime$ exhibits a sharp upturn, while
the T in the  vicinity of which long range magnetic ordering sets in is 
inferred from the shoulder/peak 
temperatures in the $\chi \prime$ plots (Fig. 1). The
additional spin-glass transition (around 2.5 K) seen in the ac $\chi$ data
for the compositions x= 0.0 to 0.3 is not marked in the figure separately
for the sake of simplicity of the figure, particularly noting that these
phases are also anyway spin-glasses. This transition must be intrinsic to
the samples considering that the materials are essentially single phase
and future investigations may focus on the exact origin of this phase.  
It may also be added that the absolute values of T$_o$ determined from
different techniques slightly (to the extent of 0.5 K) vary, which is
attributed to some degree of ambiguity in locating the actual transition
point from different experimental techniques. Therefore, the values
plotted are representative of the transition regions.  In spite of these
deficiencies, this pictorial representation serves as an useful guide to
bring out the main points. To demonstrate the novelty of the magnetic
phase diagram of this series, we compare our findings with that for a
well-studied system, CeCu$_{1-x}$Ni$_x$ (compare with Fig. 3 in Ref. 2): 
(1) The magnetic phase diagram is
reversed for both these two series of alloys, CeCu$_{1-x}$Ni$_x$ and the
present one.  (2) The ones that exhibit spin glass behavior in
our solid solution fall in the low coupling regime (that is, close to or
below the maximum of the Doniach's magnetic phase diagram as inferred from
the $x$ dependence of T$_o$), whereas the compositions close to QCP are
not spin glasses. This observation is in sharp contrast to expectations
based on previous knowledge in the field.  (3) In some of the alloys with
more than one magnetic transition, that is, in the range $x$= 0.1 to 0.3,
long range magnetic ordering precedes spin glass freezing as T is lowered,
in sharp contrast to the reverse behavior observed in the Ce-Cu-Ni series.  
To our knowledge, this kind of magnetic phase diagram has not been
reported for any Kondo alloy in the past.

We like to add that, in the neutron diffraction
data\cite{13} at 2 K for all compositions classified here as spin glasses,
no additional sharp lines appear when compared with the data at 10 K.
However, for $x$= 0.6, distinct evidence for long range magnetic ordering
is found in the neutron data. These neutron data thus endorse the main 
findings (summarised below) in this article.

To conclude, we have identified a Ce based solid solution in which
increasing chemical disorder surprisingly suppresses spin glass behavior
in favor of long range magnetic order as one moves towards Kondo-dominated state.  
 It is intriguing to note that the disorder-induced
spin-glass behavior is realised only at the weak coupling limit, but not
close to QCP, in contrast to the hitherto observed, predicted or assumed
behavior.\cite{1,2,3,4,5} Absence of this disorder effect at QCP is surprising, particularly
considering that a stoichiometric end of the solid solution is by itself a spin-glass. 
It is therefore a real theoretical challenge to understand
this unusual interplay among the disorder, spin-glass ordering and the
Kondo effect.

One of us (EVS) would like thank C. Laubschat for a visit to Dresden under
SFB463, during which this collaboration with Augsburg group was initiated.
He also thanks Sarojini Damodaran Foundation for supporting the visit of
Subham Majumdar to TU Vienna.
Research in Vienna is supported by the Austrian FWF, project P 12899.  We
thank A. Szytula for bringing us to our attention the results of his
neutron diffraction measurements before publication. We thank Kartik Iyer
for his help in sample characterization and magnetization measurements. 


\vskip0.5cm
\begin{figure}
\vskip0.5cm
\caption{Real ($\chi \prime$) and imaginary ($\chi \prime \prime $) part
of ac susceptibility as a function of temperature (logarithmic scale)
below 12 K for the alloys, Ce$_2$Au$_{1-x}$Co$_x$Si$_3$, at various
frequencies. Since, for x= 0.8, the data at various fequencies overlap,
and hence we show the data only at one frequency (1 Hz).}
\vskip0.7cm
\end{figure}

\vskip0.5cm
\begin{figure}
\vskip0.5cm
\caption{(a) Isothermal magnetization behavior of Ce$_2$Au$_{0.9}$Co$_{0.1}$Si$_3$
at selected temperatures; (b) Heat capacity as a function of temperature below 5 K for the
alloy, Ce$_2$Au$_{0.2}$Co$_{0.8}$Si$_3$. }
\vskip0.7cm
\end{figure}

\vskip0.5cm
\begin{figure}
\vskip0.5cm
\caption{Schematic representation of the magnetic phase diagram of the
alloy series, Ce$_2$Au$_{1-x}$Co$_x$Si$_3$. T$_o$ represents magnetic
transition temperature. The continuous lines through the data points serve
as a guide to the eyes.}
\vskip0.7cm
\end{figure}


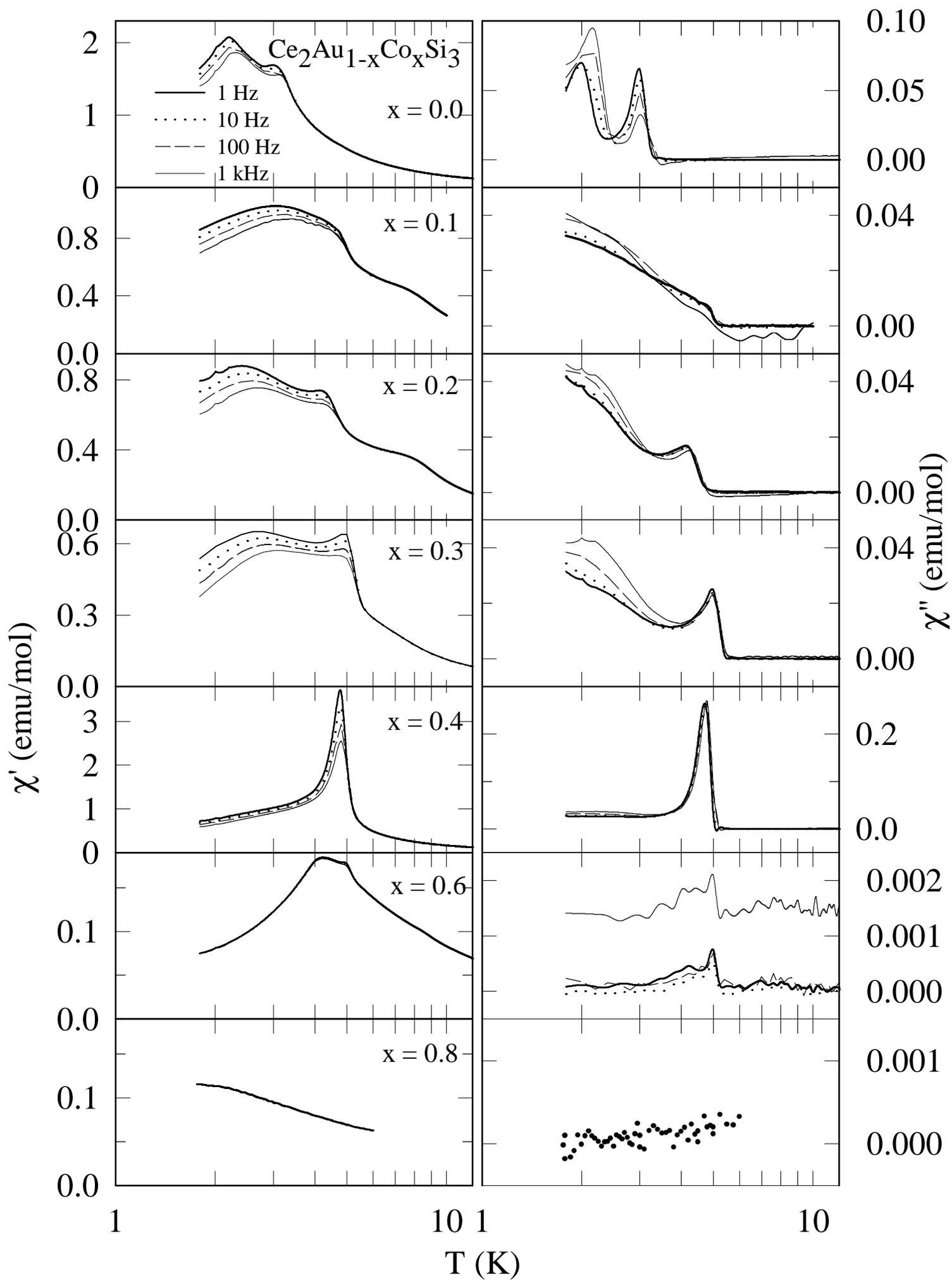
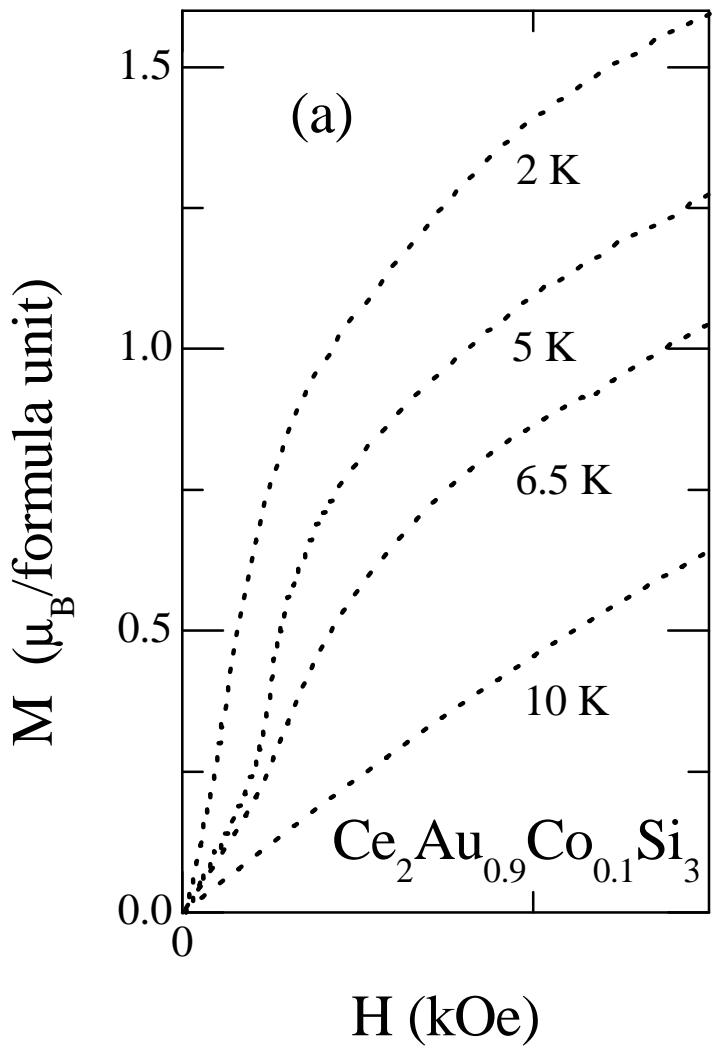
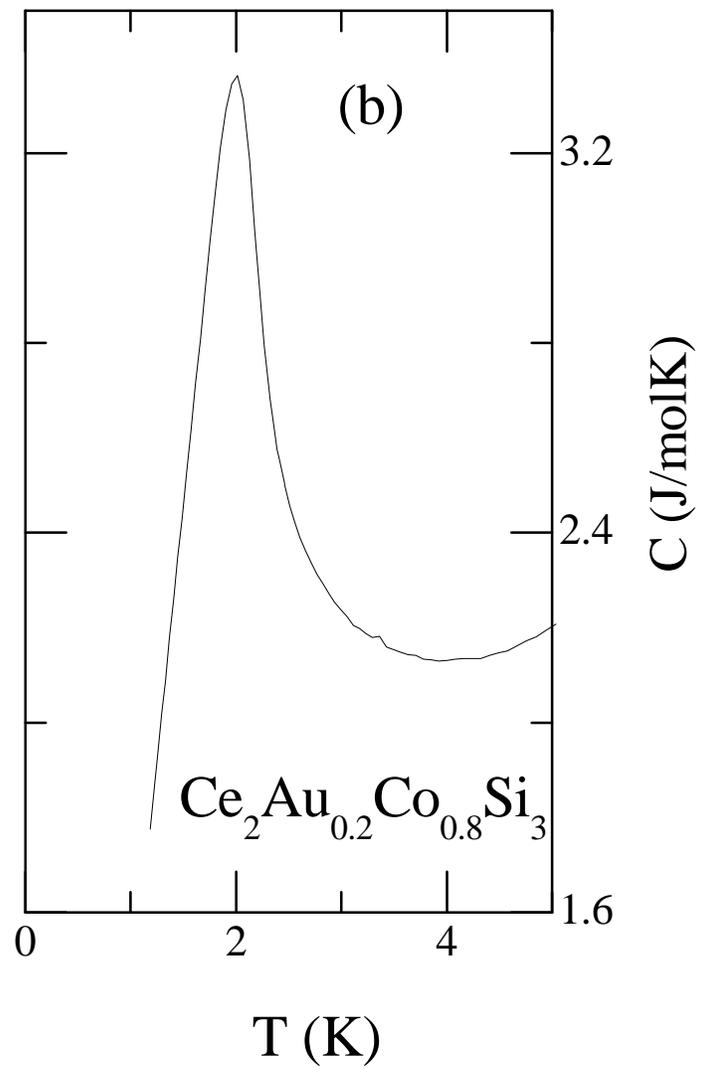
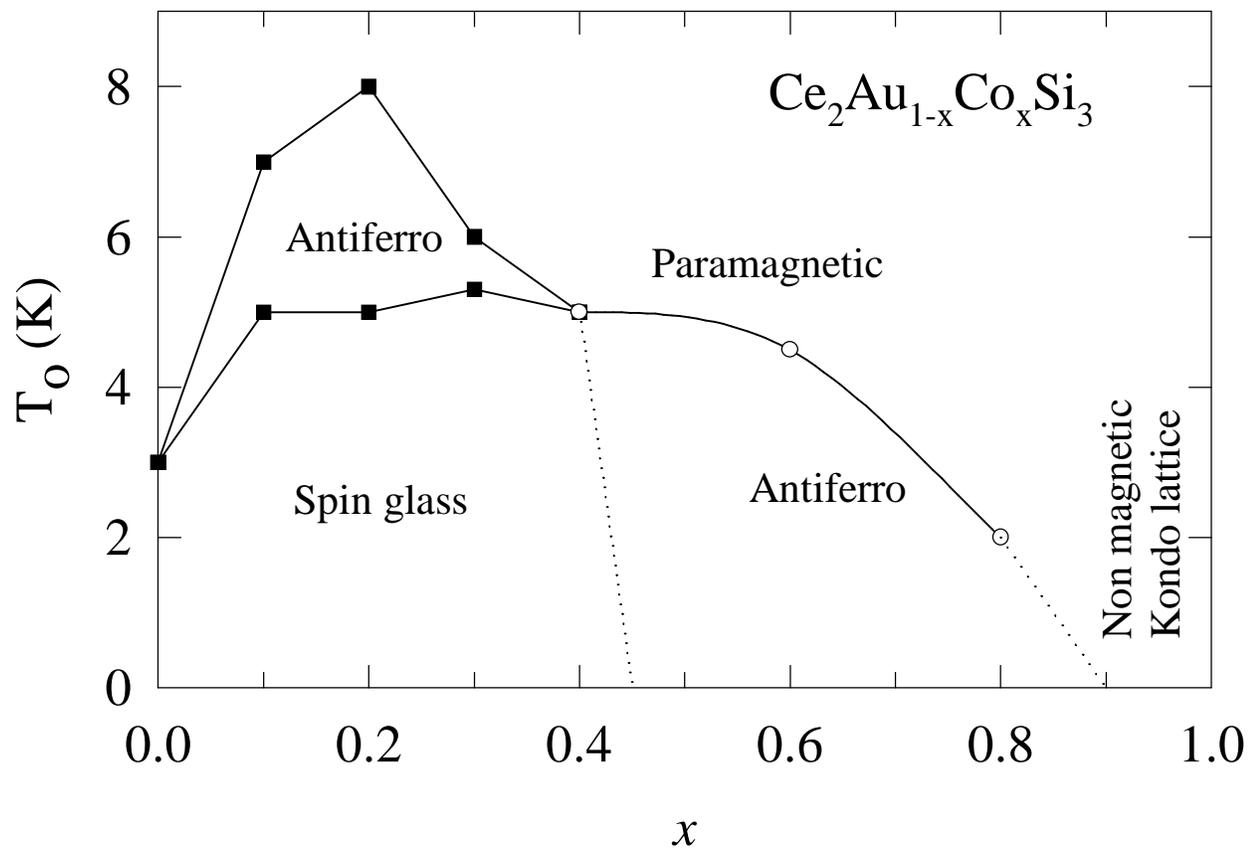
\end{document}